\def\hc{N$_{\rm H~I}$}

\documentstyle[aaspp4]{article}

\tighten
\begin{document}

\title{Global Models of the Galactic Interstellar Medium: Comparison to X-Ray and \ion{H}{1} Observations}
\author{Alexander Rosen\altaffilmark{1,2}, rosen@eclipse.astr.ua.edu}
\author{Joel N. Bregman\altaffilmark{1}, jbregman@astro.lsa.umich.edu}
\author{and Daniel D.\ Kelson\altaffilmark{3}, kelson@ucolick.org}

\received{June 15, 1995}
\revised{February 21,1996}
\accepted{May 10, 1996}

\slugcomment{Submitted to The Astrophysical Journal}

\altaffiltext{1}{Department of Astronomy, University of Michigan, Ann Arbor, MI 41809}
\altaffiltext{2}{Current address: Department of Physics and Astronomy, University of Alabama, AL, 35487}
\altaffiltext{3}{Lick Observatory, University of California at Santa Cruz, Santa Cruz,
CA 95064}

\vskip -0.5 in

\begin{abstract}

In a previous paper, we calculated numerical hydrodynamic models of the
interstellar medium in the Galaxy, which suggested that hot gas (T
$\ge$ 3 $\times$ 10$^5$ K) has a filling factor near 50\% in the
midplane, and that it is separated by cooler material (\cite{ros95}).
Here we extend the work to examine the X-ray emission characteristics
of the best model and calculate a variety of observable measures for
comparison with the observed soft X-ray background.  For five observer
locations in the disk (three hot bubbles, a cooler bubble, and a
neutral gas region), we calculate the X-ray intensities, spectra, and
hardness ratios in 0.16 -- 0.28 keV and 0.53 -- 0.87 keV bands as
Galactic latitude and cool gas column are varied.  We compare these to
strip scans of observational data, \hc\ from Dickey \& Lockman 1990,
and C band and M$_1$ band X-ray data from the Wisconsin surveys
(\cite{mcc83}).

The calculated neutral hydrogen column density distribution has a broad
range and a median value that is typically 2.5-6 times smaller than the
mean value (seen from a location in the disk or perpendicular to the
disk).  This difference between the mean and median \hc\ offers a
natural explanation for the observed difference (a factor of three)
between the average \hc\ at the solar circle and the local value
deduced from high-latitude observations.  The observed distribution of
\hc\ is similar to that seen from one of the simulated bubbles, with
the important exception that the minimum hydrogen column in the model
is too low.  The low minimum hydrogen column is a common result of the
models and indicates that neutral gas is too easily compressed into
small structures.

The models suggest that the X-ray emission in the 0.16 -- 0.28 keV band
is dominated by hot gas within 0.1 -- 0.5 kpc while in the 0.53 -- 0.87
keV band, nearly all emission originates within 2 kpc of the observer,
and often much closer.  The model X-ray emission generally hardens
toward the plane, for observers in bubbles.  Also, there are clear
examples of anticorrelations between \ion{H}{1} and X-ray emission as
well as correlations between \ion{H}{1} and X-rays, which is caused by
an increased emission measure as a shock enters a cool gas region.
Statistically, anticorrelations are slightly more common than
correlations.  X-ray spectra are calculated from the models and these
reveal that for observations to have strong diagnostic power in probing
the hot ISM, a spectral resolution of E/$\Delta$E $>$ 30 is required.

The X-ray observations reveal shortcomings in the models in that the
angular distributions of the model X-ray intensities and the hardness
ratios vary far more than the observations in either energy band.
Also, the model X-ray intensities are typically fainter than observed.
Some of these shortcomings may have been alleviated had we chosen a
more uniform and denser hot bubble, but we suggest that this problem,
as well as the low \ion{H}{1} values, might be solved by including
magnetic fields in future simulations.

\end{abstract}


\lefthead{Rosen, Bregman, \& Kelson} \righthead{Diffuse X-Ray Emission
in Simulated ISM}

\section{Introduction}

The interstellar medium of the Galaxy is very diverse, with gas
components of temperatures ranging from 10 -- 10$^7$ K in close spatial
proximity.  Each component is important either from mass
considerations, where the cold molecular gas and neutral atomic gas
account for most of the gaseous mass, or from volumetric
considerations, where the ionized phases may occupy most of the
volume.  Furthermore, the ISM in the Milky Way appears to be typical of
other spiral galaxies, where these various gaseous components are
commonly observed.

There is general understanding that the structure of our ISM is
descended from the interplay between star formation, stellar evolution
(including the essential heating events) and the gaseous disk in which
they are embedded.  Analytical models for the global ISM have focused
on issues of pressures and filling factors, usually making assumptions
about the geometry and structure of the gaseous components (e.g.,
neutral gas lies in clouds; SNR expand into a uniform medium).  Such
assumptions can be avoided in numerical models, but these calculations
are challenging even with modern computers, so it is often necessary to
examine only a piece of the problem (e.g., magnetically expanding
superbubbles) or to make other concessions when considering the global
problem.

We and collaborators have taken the path of developing global
numerical models for the interstellar medium on scales larger than the
biggest individual elements (superbubbles) so that the full range of
interstellar phenomena can develop with a minimum of assumptions.  In
this model, the stars and the gas are represented as separate fluids
that can interact.  Cool gas forms into stars at a rate
proportional to the gas density, and evolving stars return gas to the
ISM.  The stars heat the gas through a combination of stellar winds and
supernovae, with the latter being the more important of the two heating
mechanisms.  The gas evolves according to the Euler equations, modified
by these source and sink terms, as well as by optically thin radiative
cooling.

The early two-dimensional calculations with this model were for
the midplane of a galaxy (no gravity), used a first-order accurate
numerical code (the ``beam scheme": \cite{chi85}, \cite{chi88}),
and revealed a common structure for the gas, with cold dense filaments
(probably sheets in three dimensions) frequently surrounding bubbles of warm 
dilute gas.  Subsequently the model was improved by using a
second-order accurate numerical code (``Zeus-2D": \cite{sto92};
\cite{ros93}; and \cite{ros95}, hereafter RB) and two-dimensional
calculations were performed where one dimension is perpendicular to the
plane.  The models given in RB are the most realistic to date and they
include supernova heating as well as stellar wind heating.  These
models demonstrate that by matching the models to certain basic
observations, such as the scale heights of the gaseous components and
the pressures, it is possible to determine the volume filling factors
and topology of these components.

We are encouraged by the results of these models and their ability
to reproduce certain basic observations.  However, we are aware that
the model presently lacks certain important physics (most notably, a
magnetic field and a third independent dimension) and that there may be
other observables that could provide much greater diagnostic power in
identifying the model most representative of the Milky Way.  In this
paper, we compare our most successful model (from RB) to the spatial
distributions of 21 cm emission from \ion{H}{1} and of the soft X-ray
background from the Galaxy.  These observables prove to be extremely
powerful diagnostics in identifying the weakness and strengths of the
calculations and point the way toward future development.

In this paper, we will discuss the hydrodynamical model of RB in a bit
more depth in \S 2, compare the distribution of \hc\ from a variety of
positions within the model with observational data in \S 3, and compare
the X-ray intensities from the model with observations in \S 4.  Our
results are summarized and a plan for our future work in regard to
diffuse Galactic X-ray emission is discussed in \S 5.

\section{Our Hydrodynamical Simulation}

All of the simulations in RB were run with a two-dimensional version of
the Zeus hydrodynamical code, modified to allow for the evolution of
two interacting cospatial fluids.  The two cospatial fluids represent
the stars and gas in the interstellar medium, with the stellar ``fluid"
having characteristics of a mean Population I star.  The star formation
rate and mass loss rate are equal, and they are set so that their
inverse gives an effective stellar lifetime of 10$^8$ yr.  One of the
dimensions represents the vertical axis of the Galaxy, with a
time-invariant external potential imposed along this axis, which we
will designate as $z$.  The other dimension is in the plane, although
it is neither radial nor azimuthal, since the simulations in RB did not
include rotational effects, and we designate this axis as $x$.   The
simulation was run for 300 Myr, and the image chosen was at a program
time of 215 Myr.  The grid used for the simulation is a 200 $\times$
400 zone grid, and is linearly scaled with 10 pc zones in both
directions in the inner 200 rows (a 2 $\times$ 2 kpc region).  The
outer 100 rows on both the top and bottom of the grid are 2 kpc across,
but extend from 1 -- 15 kpc with the zone size increasing geometrically
along $z$ (see Figure 1a).  Also, the boundary conditions are periodic
along the vertical edges of the grid (as shown in Figure 1), and allow
outflow along the horizontal edges.

\placefigure{fig1}

In order to simplify a complex problem in the simulations of RB, the
ionization of the gas fluid was incorporated in only two aspects of the
hydrodynamical code.  Ionization effects were only crudely taken into
account in the cooling function, which was a series of power laws that
match fully ionized gas above 8000 K and gas that became less ionized
with lower temperature.  The second place where an ionization was
assumed for the gas was in the constant mean molecular weight (= 0.6)
in the calculation of gas temperatures.

Of the six simulations in RB, which varied in the rate and form of
energy injection, we analyze an image from the simulation that seems
most similar to the Galaxy, i.e.\ one that has a Galactic energy
injection rate and that includes supernovae.  This simulation naturally
recreates a multi-phase medium with cold, dense filaments (T $\sim$ 300
K, $n \sim$ 10 cm$^{-3}$) surrounding bubbles of hot gas (T $\sim$
10$^6$ K, $n \sim$ 0.001 cm$^{-3}$ in the midplane), which is
similar to the topology in Figure 9a of \cite{bur93} of the Eridanus
region.  Also, this simulation reproduces quite well the scale heights
and central densities of a three phase medium with multiple components
in each phase.

In this paper, we will define cold gas in the simulations as gas with
T, the gas temperature, $<$ 4000 K, warm gas with 4000 K $<$ T $<$ and
300,000 K, cool gas with T $<$ 300,000 K (both cold and warm gas), and
hot gas with T $>$ 300,000 K.  The upper temperature limit for the cold
gas is consistent with observations of the warm neutral medium reported
in \cite{ver94}, therefore cold gas in the simulations of RB should be
neutral hydrogen.

\placetable{tbl1}

Previously, we compared the results of the simulations to the
observed scale heights and pressures of the interstellar medium and met
with some success.  However, the observations contain considerably more
information, such as the range of intensities as seen from the Sun and
the angular distribution of these intensities.  Here we calculate the
angular distribution of \ion{H}{1} and X-ray intensities from five
different locations in the simulations: three hot bubbles, a large
region filled with both warm and hot gas, and a dense neutral region.
The temperature distributions of the five regions are shown in Figure 1.  
Some properties of the
regions are characterized in Table 1, which gives the zone locations
($x, z$), height above the midplane, mean density, mean temperature,
and the median pressure for the five regions shown in Figure 1c.

Bubbles 1 -- 3 are dominated locally by hot gas and so should be
most similar to the hot bubble around the Sun.  These bubbles were
chosen as characteristic of the simulations rather than being a close
match to our local bubble, so in the comparison that we will make
below, we will emphasize generic similarities or differences.  The
three hot bubbles from the simulation represent a range of
temperatures, densities, and heights above the midplane.

The main difference between the three hot bubbles is the average
temperature in the bubble, which is larger in Bubbles 2 and 3 than in
Bubble 1.  Variations in the average temperature in the bubbles are
caused by different elapsed times since the most recent supernova in
each bubble.  Note that Bubble 2 is also farther removed from the
midplane ($|z|$ = 230 pc, as opposed to Bubble 1, which is within 10 pc
of the midplane, and Bubble 3 with $|z|$ = 60 pc).  The median
pressures listed in Table 1 for each of the five positions show that
the hot bubbles have the highest median pressure, with the cooler gas
zones at a median pressure that is half an order of magnitude lower.
This is consistent with an observational result (\cite{bow95}) and one
result in RB, that the hot gas is overpressured with respect to the
rest of the gas.

Sample lines of sight out to a distance of 5 kpc from Position 1 are
also shown in Figure 1a.  Since only the hot gas contributes to the
X-ray intensity (see below), we display only hot gas in Figure 1b.  The
extents of the ``local" bubbles at the first three positions are shown
in Figure 1c, where we define local as the extent of each line of sight
until it first encounters a cool temperature zone.  This definition
emphasizes the X-ray brightness in the definition of the bubble's
extent.

\section{A Comparison with Galactic H I}

There are a variety of 21 cm observations of neutral hydrogen that
serve to characterize the large-scale radial and vertical
distributions, the pressure, and the angular distribution on the sky.
The Galactic \ion{H}{1} vertical distribution and pressure were
reproduced by Run E in our simulations (above), and here we extend the
comparison between model and observational data.  Comparisons between
the model and the observations are performed for the column density
distribution, primarily as a function of Galactic latitude.  Aside from
examining the \ion{H}{1} distribution, we examine the relationship
between neutral hydrogen and X-ray emitting gas.  In the soft X-ray
bands, neutral gas is the dominant absorber of X-rays and there are a
number of observations that stress this comparison.  The relationship
between the \ion{H}{1} and X-ray emitting gas is discussed in detail in
\S 4.6.

The \ion{H}{1} data (kindly provided by Lockman; see \cite{dic90}) have
been compiled from many surveys and averaged over 1\arcdeg\ $\times$
1\arcdeg.  From these data, we extracted a single strip that passes
through both Galactic Poles, and through 0\arcdeg\ latitude at
longitudes of 90\arcdeg\ and 270\arcdeg.  Two considerations assisted
in the choice of this strip: 1) that the simulations of RB did not
include rotational effects, and 2) that we wanted to compare our model
with a symmetrical distribution of \ion{H}{1} in the Galactic disk,
i.e. one of equal length through the Galactic disk at both longitudes.

From our model, we calculate a distribution of cold and cool gas as a
function of latitude for each of the five positions.  There can be only
one such distribution per location since the simulation has only two
independent coordinates.  The column density is calculated from the
lower left-hand corner of the zone location (Table 1) and along a line
with a constant 1 cm$^2$ cross-section.  We have computed these column
densities for all latitudes with 0\fdg5 separation (720 lines of
sight), and thereby generated a simulated strip scan.  For each line of
sight, we have calculated the column density to a distance of 5 kpc
from the initial position, making use of the periodic boundary
conditions employed in the (2 kpc wide) hydrodynamical simulation where
necessary.  The bending of sample lines of sights, shown in Figure 1a,
once out of the central region (rows 100 -- 300) is caused by the
logarithmic scaling of the vertical size (from a height of 1 kpc 100
rows from the midplane to 15 kpc at the edge of the grid) of zones in
these rows.  We have not smoothed the variations of any of our strip
surveys in angular resolution.

\placefigure{fig2}

The maxima in the \hc\ plots always occur near the midplane ({\it b} =
0\arcdeg L or 0\arcdeg R, see Fig.\ 2), which results from the line of
sight crossing the entire width of the grid 2.5 times through the
densest gas.  The minima in \hc\ are usually, although not always, at
lines of sight near each pole at {\it b} = $\pm$ 90\arcdeg; an
exception to this is a minimum in Bubble 1 at {\it b} = 30\arcdeg L.

\placetable{tbl2}

First, we compare the distribution of column densities between the
model and the observations.  To do so, we calculated the minimum, low
quartile (25\%), median, high quartile (75\%), maximum, mean, and the
fluctuation statistic, $\delta_i = {\sqrt{\langle i^2 \rangle - \langle
i \rangle^2} \over \langle i \rangle}$, in the strip surveys in
\hc\ (see Table 2).  The fluctuation statistic, listed in the last
column of this table, is a measure of the width of the distribution.
In Table 2, we have also included similar statistical quantities for
the single strip of data from the DL dataset.  The \ion{H}{1}
statistics for the model data contain many trends, all of which can be
explained by a combination of two effects: that smaller values of
\hc\ depend on the local conditions, while the larger values of
\hc\ (specifically the maxima) depend primarily on the height of the
position above the midplane ($z$ = 200 zones).

The statistical quantities for lower values of \hc\ are determined by
the local environment; specifically, the minima (column 2), low
quartiles (column 3), and medians (column 4) of the three bubbles are
all close to each other, while those from Position 4 are the smallest
in each column, and those from Position 5 are the largest.  Position 4
has the smallest values in these columns because the superbubble at
{\it b} = +90\arcdeg\ has heated the cold gas so that it is completely
ionized toward some latitudes in this direction.  The positions in the
three bubbles of hot gas have intermediate values in columns 2 -- 4 of
Table 2, and these positions all have log (minimum \hc) $\sim$ 18.5 and
median values of log \hc\ $\sim$ 20.1.  The largest values of
\ion{H}{1} minima, low quartiles, and medians are associated with the
dense clump of cold gas at Position 5.

The positional variation of the maxima (column 6 of Table 2) and the
high quartile (column 5) model data follow different patterns from the
variation of the lower statistical quantities.  The physical
significance of this difference is revealed by comparing column 6 in
Table 2 with the height above the midplane, $|z|$, which is given in
column 3 of Table 1.  The three positions with $|z| =$ 10 pc all have
roughly the same maxima (log \hc\ $\sim$ 22.4), and this is caused by
low latitude lines of sight that pass through the disk.  The smallest
maximum \hc\ occurs at Position 2, which is at a large enough height
that the lines of sight that pass through the midplane do so for a
shorter path length than sightlines closer to the midplane.

\placefigure{fig3}

With the exception of some very low \hc\ in a portion of the sky, the
distribution of \hc\ surrounding Bubble 3 is similar to the observed
Galactic distribution of \hc.  The model data from Bubble 3 is most
like the data from the DL strip in every column of Table 2 (except for
the high quartile in column 5), including the fluctuation statistic.
We display the latitude variation of the strip taken from the DL
\hc\ data with the ones created from each of the five positions in the
simulation in Figure 2.  In addition, we plot histograms of the six
distributions, for each of the 5 positions and the DL strip, in Figure
3.   Each of the simulated sets of \hc\ within Bubbles 1, 2, and 3
(Figures 2a, 2b, and 2c) has a smaller minimum value than the
observations.  From Figures 2 and 3 and Table 2, the model
distributions from Positions 4 and 5 quite definitely do not fit the
observations, because Position 4 has too many low \hc\ lines of sight
and Position 5 has too few.  Of Bubbles 1, 2, and 3, the latitude
dependence of \hc\ (Figure 2) shows that only Bubble 3 has nearly as
many lines of sight with \hc\ above the observational strip of data as
below it.  Also, the histograms in Figures 3a, 3b, and 3c show that
Bubble 3 has a distribution that is most similar to the DL data in
Figure 3f, although there are enough \hc\ sightlines with small values
seen from Bubble 3 that it is not a very good match.  By comparison,
the observed distribution of \ion{H}{1} has very few regions of low
column density, with the minimum \ion{H}{1} column density in the Ursa
Major region (log \hc\ = 19.64, \cite{jah90}) and a sharp cutoff below
log \hc\ = 20.0 (see Figure 4 in DL for the histogram of the entire
dataset).

This comparison between observations and simulations leads to an
explanation for the observational fact that the \ion{H}{1} column above
the Sun is about one-third the mean value at the solar circle.  This
difference in the mean and local columns either is an important clue to
the nature of the neutral ISM, or it is an unrepresentative situation
that occurred by chance.  Our calculations suggest the former.  An
inspection of mean and median \hc\ in the model data reveals that the
median is consistently lower than the mean, and for the bubbles of hot
gas the ratio of the two is close to the observed value.  The ratio of
mean to median in columns 4 and 7 of Table 2 is 2.5 -- 6 in each of the
positions within bubbles of hot gas.  This is similar to the difference
between the typical column of neutral hydrogen at high latitudes
($\sim$ 1 $\times$ 10$^{20}$ cm$^{-2}$) and the half-column through the
observationally derived mean distribution of the entire set of
\hc\ data (from DL, $\sim$ 3 $\times$ 10$^{20}$ cm$^{-2}$).  This
result occurs because of the low filling factor of the \ion{H}{1}, so
only a few \ion{H}{1} structures are found along any line of sight
through the disk, and some of these structures contain considerable
\ion{H}{1} mass.  Also, this leads to the prediction that for lines of
sight through the \ion{H}{1} disks of external galaxies, the \ion{H}{1}
absorption column toward a point source will usually be lower than the
\ion{H}{1} emission column seen in a large beam (e.g., a beamsize
greater than the characteristic structure size of a few hundred pc).
This has the implication that in external galaxies the determination of
the \ion{H}{1} spin temperature will typically be underestimated from
standard techniques (e.g., the ratio of 21 cm absorption to emission
measures; for the above beamsize situation).

\section{A Comparison with the Galactic Soft X-Ray Background}

The soft X-ray observations can provide some of the strongest
constraints for any model of the ISM, although the use of this data is
complicated by the presence of an extragalactic X-ray background and by
the absorption effects of cool gas in the Galaxy.  Here, we present
calculations of the X-ray emission properties of our models which are
compared to X-ray observations as well as to the absorbing \ion{H}{1}
material.

\subsection{An Overview of X-Ray Background Observations}

The X-ray background is dominated by extragalactic sources at
energies above about 1 -- 2 keV, but at lower energies, E $<$ 1
keV, the X-ray background is a local phenomenon that is caused by hot
gas in the Galaxy (reviews by \cite{mcc90}; \cite{fab92}).  The
emission in the softest bands (0.1 -- 0.5 keV) is dominated by the
Local Bubble of hot gas surrounding the Sun, which is 100 pc in size
and has a temperature near 1 $\times$ 10$^6$ K.  The absorption by
neutral Galactic gas prevents us from sampling far into the Galaxy at
the lowest energies, 0.1 -- 0.2 keV, where an optical depth of unity is
reached after only 0.2 -- 1 $\times$ 10$^{20}$ cm$^{-2}$.  Because the
absorption cross-section decreases quickly with energy (e.g., Morrison
\& McCammon 1983), an optical depth of unity is reached for a path
length in the disk of about 1 kpc at an energy of 0.7 keV and 5 kpc at
an energy of 1.5 keV.  At these energies, it is possible to observe
features beyond our Local Bubble.  Consequently, we will examine the
X-ray background at two energy bands, near 0.2 keV and 0.7 keV.

Sensitive large-scale surveys of the soft (Galactic) X-ray
background were carried out by the Wisconsin X-ray group with
sounding rockets (7\arcdeg\ resolution; McCammon et al. 1983), from the
SAS 3 satellite (0.25 keV map with 4.5\arcdeg\ resolution;
\cite{mar84}), and with the A2 LED detectors on HEAO 1 (0.12 -- 3 keV
and 3\arcdeg\ resolution; 
\cite{gar92}).  These observations showed a
structured X-ray sky, with order-of-magnitude variations in the diffuse
emission, and brightening in the softest X-ray bands toward the north
and south Galactic poles.

The X-ray satellite ROSAT improved the ability to study of the
X-ray background due to substantially better angular resolution
(1\arcmin), which allowed high sensitivity shadowing experiments to be
performed (review by Burrows \& Mendenhall 1994).  In these shadowing
experiments, one observes in the direction of neutral gas clouds
(\ion{H}{1}) with a known distance that places them beyond the Local
Bubble.  If all of the soft emission is from the Local Bubble, no
shadow is seen, so it was extremely exciting that shadows were seen in
some directions (\cite{bur91}), usually along lines of sight out of the
plane.  In the plane, shadowing by clouds at energies near
\slantfrac{1}{4} keV is uncommon.  At higher energies (e.g.,
\slantfrac{3}{4} keV), observations in the plane of the Galaxy commonly
reveal shadowing, indicating that most of this emission is from within
a few kiloparsecs of the Sun (Burrows \& Mendenhall 1994), and
\cite{sta94} present a model for this distribution (note that at low
latitude, the disk blocks the extragalactic contribution to this
component).

In a complementary effort, workers have sought to identify
individual features or structures, and have found several large
bubbles, which are explained as reheated supernova remnants and
superbubbles.  Some examples are the Cygnus superbubble (\cite{cas80}),
the Monogem Ring (\cite{plu94}), and the Orion-Eridanus enhancement
(\cite{nou82}), all nearby objects (few hundred pc) that subtend large
angles on the sky (several tens of degrees).  The study of the hot ISM
at kiloparsec distances requires angular resolution of 0.3 --
1\arcdeg\ at 0.5 -- 1.5 keV energies, and some programs are being
carried out with the ROSAT PSPC All-Sky Survey (\cite{sno94a};
\cite{sno95}; and \cite{sno96}).

\subsection{The X-Ray Observations Used Here}

As in the comparison of \ion{H}{1} data, we require characteristic
measures of intensity as a function of latitude around the sky.
For these all-sky X-ray scan data, we use the observations obtained by
the Wisconsin group in their C and M$_1$ bands, which have 20\%
response points defining the bands at 0.16 -- 0.284 keV and 0.44 --
0.93, respectively (McCammon et al.\ 1983).  These observations have
angular resolutions of 7\arcdeg\ (FWHM) in C band and 6.2\arcdeg\ in
M$_1$ band, and although that is considerably inferior to ROSAT
(effective angular resolution of about 1\arcdeg\ in the ROSAT All Sky
Survey, due to the need to bin the data to improve photon statistics),
the ROSAT data were not publicly available at this writing and their
equivalent energy bands are not as cleanly defined (nevertheless, a
qualitative comparison to some of the ROSAT observations is made).

The X-ray surface brightness as a function of Galactic latitude
was extracted for a strip that passes through the Galactic longitudes of 90\arcdeg and 270\arcdeg.  There are a few particularly bright portions 
of the X-ray data due to discrete sources and they have been removed (3 
of 360 points in the C band strip and 11 of 360 points in the M$_1$ 
band strip).

\subsection{The X-Ray Model Intensities}

For each of the five positions from which we generated the neutral
hydrogen strip scans, we have created X-ray intensity strip surveys
(with the same angular separation between adjacent lines of sight,
0\fdg5) in two soft X-ray bands: a band between 0.155 -- 0.284 keV,
which is similar to the Wisconsin C band (0.16 -- 0.284 keV;
\cite{mcc83}) and also the ROSAT R2 band (the 10\% points are 0.14 --
0.284 keV; \cite{sno94b}), and one between 0.532 -- 0.873 keV, which is
similar to but narrower than both the Wisconsin M$_1$ band (0.44 --
0.93 keV) and the ROSAT R4 band (0.44 -- 1.01 keV).  Our energy bands
do not include an instrumental response as a function of energy, which
is different for each instrument.  The energy bands that we chose are
two of the six energy bands for which \cite{ray76} have discussed the
cooling coefficient, $\Lambda$, as a function of temperature.  We will
refer to each model band by its mean (0.22 keV and 0.70 keV), and these
values will always indicate simulated data.

The computation of the X-ray intensity (I) is based on a combination of
the program outlined in \S 3 and on an updated version of the X-ray
emission code described in \cite{ray77}.  The cosmic abundances used
are from \cite{all73} (in the log relative to hydrogen = 12.00):  He,
10.93; C, 8.52; N, 7.96; O, 8.82; Ne, 7.92; Mg, 7.42; Si, 7.52; S,
7.20; Ar, 6.90; Ca, 6.30; Fe, 7.60; and Ni, 6.30.  This code was used
to generate a series of spectra at different temperatures and for each
band.  The temperatures of the spectra were between log T = 5.5 to 6.7,
with temperature steps of 0.05 dex (25 different temperature values).
For each band, a single spectrum is composed of 256 points (bins) with
an energy separation of 0.50 eV and 1.33 eV across in the 0.22 keV band
and 0.70 keV band, respectively.  We calculate the contribution to the
X-ray intensity in each zone of hot gas by multiplying the emissivity
from the most appropriate of the tabulated spectra (as determined by
the temperature of the zone) by the emission measure (the product of
gas density squared and length of the sightline across the zone).

The radiative transfer along the line of sight is simplified because
the emitting plasma has a low optical depth and the absorbing material
neither scatters nor emits X-rays.  Therefore, along a line of sight,
radiative transfer amounts to discrete additions of source spectra or
discrete absorption of the net incoming spectral energy distribution.
We assume that absorption is produced by gas below temperatures of
300,000 K, although nearly all of the attenuation is due to much cooler
material by virtue of greater column densities.  The opacity is
calculated at the central energy for each of the 256 {\it bins} within
each band, and the path length of the sightline across the zone.  For
the absorption cross-section, we used the \cite{mor83} prescription,
which is of the form (c$_0$ + c$_1$E + c$_2$E$^2$)E$^{-3} \times
10^{-24}$ cm$^2$, where c$_0$, c$_1$, and c$_2$ are the coefficients of
an analytic fit and E is in keV.  The calculations of the intensity are
for a Galactic line-of-sight that begins 5 kpc from the observer
locations; at these energies, Galactic contributions from beyond 5 kpc
are very small.

\placefigure{fig4}

The effects of an extragalactic X-ray background contribution
(henceforth, XRB) to the calculated X-ray intensities are important,
especially in the higher energy band and for directions of low
absorbing column.  For each line of sight at each position, we have
added an intensity equal to E$_0$ e$^{-{\sigma_{\rm eff}}{\rm N_{\rm
cool}}}$, where E$_0$ is the constant intensity emitted by the
extragalactic background, and $\sigma_{\rm eff}$ is the effective
cross-section (in units of cm$^2$).  Unfortunately, E$_0$ is not known
{\it a priori}, but is estimated based upon an extrapolation from the
XRB at higher energies; to account for this ambiguity, we consider
multiple values of E$_0$.  Estimates based on observations suggest that
the XRB at \slantfrac{3}{4} keV could be as much as 50 -- 65\% of the
observed flux in some directions (\cite{bur94}). Therefore, for E$_0$
we used values of double and triple the median of the I$_{0.70~keV}$
(without an XRB) medians for the five strip surveys, or E$_0$ = 1.2 and
1.8 $\times$ 10$^{-5}$ (in our intensity units).  Assuming an E$^{-1}$
photon spectrum for the XRB, the flux-weighted mean energy for the XRB
in the 0.532 -- 0.873 keV band is 0.688 keV, and we use this energy to
calculate a cross-section.   Then, we compute the resultant additions
for each value of E$_0$ to the 0.70 keV intensity with the cool gas (T
$<$ 300,000 K) column density at each line of sight, and plot these in
Figure 4.  Also, we have plotted the effect an XRB would have on the
0.22 keV energy band for a single value of E$_0$ in Figure 4.  The mean
energy in the 0.155 -- 0.284 keV band with an E$^{-1}$ spectrum is
0.213 keV, and we used a value of E$_0$ = 4.8 $\times$ 10$^{-5}$, which
is three times the larger of the two E$_0$ used in the 0.70 keV band
(and is consistent with an E$^{-1}$ spectrum in a band with $\onethird$
the energy).

A stellar contribution to the diffuse Galactic X-ray intensity is not
included in this analysis.  The coronae of normal stars (e.g., dM
stars) may produce as much as 10 -- 20\% of the diffuse Galactic X-ray
emission in the 0.28 -- 1.0 keV band (\cite{cai86}), but $\lesssim$ 3\%
in the 0.15 -- 0.28 keV band (\cite{ros81}).  Also, since stellar point
sources are subtracted from recent X-ray surveys, such as the ROSAT
studies of diffuse Galactic emission, we believe that this omission of
a stellar contribution will not have strong consequences for our
analysis.

\placetable{tbl3}

\placetable{tbl4}

\placetable{tbl5}

The resulting X-ray intensities are given in units of 2.078 $\times$
10$^{-12}$ erg cm$^{-2}$ s$^{-1}$ arcmin$^{-2}$ (Fig.\ 4). 
This intensity corresponds
to $\sim$ 1 ct s$^{-1}$ arcmin$^{-2}$ over the entire 0.1--2.4 keV ROSAT band (Snowden et al.\ 1994b).  Some sample conversions between our intensity
and the cts s$^{-1}$ arcmin$^{-2}$ received by ROSAT, using the on-axis
energy response function, are given in Table 3 in each of the R2 and R4 bands.   We also include in Table 3 a
conversion of our intensity to cts s$^{-1}$ in the Wisconsin C and
M$_1$ band, from the response to an E$^{-1}$ spectrum in Table 3
(although the response is not very sensitive to the spectral index of
the incident photons) in McCammon et al.\ (1983).  These conversions
are needed for the data in all of the figures with X-ray intensities
and also in Tables 4 and 5, where we list the same statistical
quantities for each X-ray band (with an XRB) as we did in Table 2 for
the \hc\ data.

\subsection{Contributions to the X-Ray Emission as a Function of Depth}

One of the important uses of the simulations is to better understand
the contributors to the X-ray emission as a function of distance from
the observer.  In general, we find that the ISM within 1 kpc of the
observer contributes $\gtrsim$ 90\% of the resultant X-ray intensity,
at 0.22 keV and 0.70 keV for both midplane and high-latitude lines of
sight (see Figure 5, which is a plot of cumulative X-ray intensity
vs.\ distance and should be ``read" starting from the right side of
each panel).  The two lines of sight displayed in Figure 5 illustrate
the inhomogeneous nature of the ISM, with one line of sight from Bubble
2 in the direction of Bubble 3 (at a latitude of -33\arcdeg R from a
line parallel to the midplane, note that for Position 2, {\it z} = 230
pc), and the other from Position 4 at a latitude of 1\arcdeg L.   From
the position within Bubble 2 (see Figure 5a), the ISM at a distance of
300 -- 700 pc contributes much of the X-ray intensity, this is where
the line of sight crosses Bubble 3.  The diminution of X-ray intensity
at 300 pc, which is significant only in the softer 0.22 keV band, is
caused by some very dense, cold gas at the boundary between Bubbles 2
and 3.
From Position 4 at $b$ = 1\arcdeg L (see Figure 5b), the line of sight
crosses Bubble 3 and enters a clump of cold gas (at $x, z$ =  142 --
145, 200) with \hc\ $\approx$ 1.5 $\times$ 10$^{21}$ cm$^{-2}$ (roughly
10\% of the total \hc\ for the line of sight).  The constant 0.70 keV
intensity within a distance of 300 pc in Figure 5b is caused by gas
that is warm, but not hot enough to contribute substantially to the
X-ray intensity.  The X-ray intensity from Bubble 3 is seen at $d$ =
300 -- 500 pc, with a decrease over 40 -- 50 pc (4 -- 5 zones) at the
closer boundary of the bubble in both X-ray bands, and the 0.22 keV
intensity is reduced to 10$^{-8}$ in these units (and is not shown in
Figure 5b).

\placefigure{fig5}

Since the grid is 2 kpc across, and the line of sight extends to 5 kpc,
features for the line of sight close to the midplane repeat, e.g.\ the
contributions of Bubble 3 are also seen at $d$ = 2400 pc and $d$ = 4400
pc in Fig.\ 5b.  Similarly, contributions from Bubble 1 are seen at $d$
= 1100 -- 1200 pc and $d$ = 3100 -- 3200 pc, and contributions from the
hot gas at the left edge of the grid (crossing the line of sight at
{\it x, z} = 4 -- 26, 201) are seen at $d$ = 1500 -- 1750 pc and $d$ =
3500 -- 3750 pc, while attenuations from near Position 5 are seen at
$d$ = 900 pc and $d$ = 2900 pc (no X-ray intensity is emitted beyond
$d$ = 4700 pc).

The two lines of sight described by Figure 5, which are typical,
indicate that at 0.70 keV, emission contributions from structures
beyond 2 kpc will be difficult to study.  At Position 4 (Fig.\ 5b), the
considerable integrated emission measure beyond 1.5 kpc contributes
less than 2\% of the observed emission, and studying a structure that
is defined by such a small fluctuation against the X-ray background is
likely to be impossible.  Other lines of sight are less favorable
(Fig.\ 5a), and at 0.22 keV, contributions beyond the local or adjacent
bubble are negligible.

Variations of the Galactic X-ray intensity in our model are caused
primarily by variation within the local bubbles, which is caused by
both the size of and the temperature variation within each bubble, but
also with a contribution of non-local (but Galactic) intensity that
varies with latitude.  Bubbles of hot gas created in the hydrodynamical
simulation naturally have non-uniform shapes and contain gas at a
variety of temperatures (see Figure 1c).  For example, the three-lobed
structure of Bubble 3 leads directly to three broad maxima in the X-ray
strip surveys (Figures 4h and 4i).  This is consistent with the success
of the displacement model (\cite{sno90}) at reproducing much of the
X-ray background at energies below 0.284 keV.  Viewing the evolution of
the simulation shows that much of the hot overpressured gas in Bubble 3
is expanding into regions of colder and denser material, similar to
what is envisioned in the displacement model.

The additional X-ray intensity from hot gas beyond each of the
simulated local bubbles has a large variation with latitude.  For
example, in Figure 4e at 0.22 keV from Bubble 2 the non-local
contribution (that can be estimated in Figures 4b, 4c, 4e, 4f, 4h, and
4i, by comparing the total intensity without an XRB, the solid line,
with the local contribution, the dashed line) varies from zero at most
latitudes to as much as 1 dex near {\it b} = 30\arcdeg L and -20\arcdeg
L.  Most of this non-local intensity is from hot gas that is near
Bubble 2 (at {\it x, z} = 80, 230), but is beyond some intervening cool
gas from the observing position within the bubble. This large variation
of the non-local intensity is consistent with an observational estimate
that the halo is eight times brighter toward Draco than toward Ursa
Major at \slantfrac{1}{4} keV (SHJLMS).  The non-local contributions in
the 0.70 keV band are often larger than those in the 0.22 keV band
for each of the three bubbles, particularly as seen from within Bubble
1 (see Figures 1b and 1c), which contains relatively cooler hot gas (T
$\sim 5 \times 10^5$ K, see Table 1) than the other two bubbles.  In
this case, the larger increment in the harder band is primarily a
consequence of the lower intensity of this cooler hot bubble in the
higher energy bands.

\subsection{The Angular Distribution of the X-Ray Emission}

From Figure 4, it is obvious that our model generates X-ray emission
that is in general both fainter than in the observations and has a
larger angular variation.  These failings suggest that we are in a
Local Bubble that is more homogeneous and has a higher emission measure
than any of the bubbles in our model (although Bubble 3 appears closest
to observations in a number of categories).  We discuss this in greater
depth in \S 5 with regard to future work.  In this section, we examine
some of the specific details of the angular variation of the simulated
X-ray emission, as well as the relative hardness in the two bands.

The angular distribution of the 0.22 keV model X-ray emission shows
many interesting details from each of the bubbles (in Figures 4b, 4e,
and 4h).  Due to the short mean free path of photons in this soft X-ray
band, the size and shape of the bubble primarily determines the nature
of the modeled X-ray emission.  In general, this means that the maximum
and minimum X-ray intensity within each strip scan can occur at any
latitude.  For example, Bubble 1 has an overall maximum near 0\arcdeg L
with a secondary peak near -60\arcdeg L; both of these are included in
the non-local emission from the large nearby bubble that subtends
latitudes between 10\arcdeg L and -90\arcdeg\ from Bubble 1.  From
Figure 4b, the minimum values seen from Bubble 1 are in the latitude
ranges 30\arcdeg L -- +90\arcdeg\ and -80\arcdeg R -- 10\arcdeg R.
These low intensities are caused by a small path length across the
bubble and a lack of hot gas beyond the Bubble 1 in these directions
(see Figure 1).

For Bubble 2, the peak in the 0.22 keV emission is at latitude
45\arcdeg L, which contains the ridge of very high temperature gas (4
-- 5 $\times$ 10$^6$ K) within Bubble 2.  Note that there is also a
broad secondary maximum 180\arcdeg\ away from this (at $b$ = -45\arcdeg
R), a result of the shorter path length through the high temperature
ridge that is wider in that direction.  The minimum for the 0.22 keV
intensity in Bubble 2 is at $b$ = -30 -- -40\arcdeg L, and this is in the
direction of some very cold (and presumably dense) gas roughly between
Bubbles 1 and 2.

The 0.22 keV intensity for Bubble 3 is brightest near one of the poles
(+90\arcdeg) and the deepest minima is near the plane (0\arcdeg L),
although the situation is reversed toward the other pole and plane
crossing; this is a consequence of the three-lobed shape of Bubble 3.
Of particular note is that this is the simulated bubble with the most
smoothly varying X-ray emission, because the observations have even
smaller fluctuations.

A statistical analysis of the angular distribution of the I$_{\rm
0.22~keV}$ emission with an added XRB is presented in Table 4, where
there are some similarities in the properties of the three bubble
locations, especially as compared to Position 4 (located in warm gas)
or Position 5 (located in cold gas).  The positional variation is
smallest in the maxima (column 6 of Table 4), and increases
monotonically as the statistic describes lower ordinal values of X-ray
intensity (ending with a largest range of values of 0.22 keV model data
in the minima --- column 2 of Table 4).  As with the \hc\ data, Bubble
3 has the closest distribution of 0.22 keV model data to that of the
observational strip from the Wisconsin C band, but the model has a much
larger range than the observations (cf.\ a $\delta_{\rm I}$ of 0.9 in
the model vs.\ 0.4 in the data).

In the model X-ray emission at the harder energy, 0.70 keV, we have
seen that contributions to the X-ray intensity come not only from the
local bubble, but from within 2 kpc of the observer.  Along this path
length, the competition between emission and absorption determines
whether the X-ray intensity has a local maximum or minimum near the
plane.  For the conditions in our simulations, the two effects are
comparable, and although we generally find a brightening near the plane
at 0.70 keV, there are clear examples of absorption effects near
the plane as well as bright regions at moderately high latitudes.  As
for the 0.22 keV emission above, we now examine the angular
distribution in the three hot bubbles.

Since Bubble 1 is a region of cooler hot gas ($\sim 5 \times$ 10$^5$
K), most of the emission at 0.70 keV is from beyond 
the local bubble at most latitudes.  The brightest regions are toward 
the plane, although
the hot gas just to the left of Bubble 2 in Figures 1a and 1b (at $x,
z$ $\approx$ 80, 230) generates a brighter spot at a latitude of
50\arcdeg R.  In general, for Bubble 1 the 0.70 keV emission from the
local bubble is so small that it is dominated by the XRB.

For Bubble 2, much of the emission at 0.70 keV is from the local
bubble, as was the case for the 0.22 keV intensity.  Non-local
contributions are usually located near the plane, $|b|$ =
0\arcdeg\ (note that this is 250 pc above the midplane in Figure 1).
Two main sources of this non-local emission are the hot gas near $x, z$
= 20, 210 and a dim version of Bubble 2 from 2 kpc away, which is a
consequence of the periodic boundary conditions.  There is some bright
0.70 keV emission at moderate latitudes, specifically at $b$ =
30\arcdeg L -- 60\arcdeg L, from the ridge of hot gas within Bubble 2
and the hot gas next to Bubble 2; -40\arcdeg L, from the hottest gas at
$x, z$ = 165, 125 in the large bubble next to Bubble 1, and -30\arcdeg
R -- -60\arcdeg R, from Bubble 3.

For Bubble 3, the 0.70 keV intensity is dominated by the hot local
bubble.  Even one of the brighter regions, toward $b$ = -50\arcdeg R --
-70\arcdeg R, that appears to be non-local emission is from the same
bubble, since the sightline from the position within the bubble crosses
the finger of cold gas at $b$ = -90\arcdeg.  Also, the three-lobed
structure that is seen at 0.22 keV is less evident, because one of the
lobes (toward $b$ = +90\arcdeg ) is at a significantly lower temperature
than the rest of the bubble.  The
largest, highest temperature lobe is the brightest region between $b$ =
-70\arcdeg R and 20\arcdeg R (see Figure 1).

The I$_{\rm 0.70~keV}$ strip scans for Bubbles 1, 2, and 3 shown in
Figures 4c, 4f, and 4i contain a dip to differing degrees in the
intensity near the midplane.  This decrease in 0.70 keV intensity is
most easily seen from Bubbles 1 and 2, and to a lesser extent from
Bubble 3.  This dip exists in observations; it is seen both in the
Wisconsin M$_1$ map (\cite{mcc83}) and in the \slantfrac{3}{4} keV map
in the ROSAT All-Sky Survey (\cite{sno96}).  While the strip scan that
we display and analyze from the Wisconsin M$_1$ data does not have a
significant dip near the midplane, the entire M$_1$ map does contain
this feature.  The strip that we are concerned with goes through the
midplane at longitudes of 90\arcdeg, which is near the Cygnus Loop, and
270\arcdeg, which has had (high brightness) data associated with the
Vela/Puppis SNR removed; therefore, the dip is not as easily detected
in the strip we chose.

The statistical quantities of the 0.70 keV model data in Table 5 are
most dependent on the spatial distribution of the hottest gas in the
simulation within approximately 2 kpc of each position.  Since the
simulated region is only 2 kpc wide and the hottest gas ($> 2 \times$
10$^6$ K) is confined to within 400 pc of the midplane, each of our
five positions is close enough to all of the hottest gas that each
position detects the same sources of 0.70 keV intensity.  As with the
0.22 keV data, the range of the model data decreases for the higher
ordinal statistics, with the maxima of the 0.70 keV model data in
Bubbles 1 -- 3 and Position 4 within $\pm$ 15\% of their mean. All of
our statistical quantities of the model 0.70 keV band intensities are
considerably smaller than their counterparts in the observational
data.  The overall dimness of the 0.70 keV model data suggests that the
hot gas in the model is cooler than the Galactic hot gas phase, which
could be remedied by a higher energy injection rate, and we discuss
this further in \S 5.

Neither of the ranges of the 0.22 keV and 0.70 keV model data in Tables
4 and 5 compare well with the ranges of the observational data in
\cite{san77} (B band, E = 0.1 -- 0.18 keV, and C band data), Burrows \&
Mendenhall (1991) (C band), SHJLMS (C band), and in the
\slantfrac{1}{4} keV and \slantfrac{3}{4} keV ROSAT All-Sky Survey maps
(Snowden et al.\ 1995b).  The large scatter in both the observations
and the model prevents a simple model of the form I = I$_{\rm 0}$ +
I$_{\rm 1}$e$^{-\sigma N_{H I}}$ from reproducing all of the data
(particularly from within a bubble), where I$_{\rm 0}$ is the local
X-ray intensity and I$_{\rm 1}$ is the X-ray intensity at some distance
removed from the observer and is completely attenuated by the entire
\ion{H}{1} column.  While the ranges in the model data are larger than
that for a simple model, they are also larger than in the
observations.  For example, there is a small ratio of maximum to
minimum of only two to five in the \slantfrac{1}{4} keV (or C band)
intensity in the observations and the smallest of the ratios from the
simulation is of order 2 dex.  Even bright supernova remnants that are
excluded from observational datasets, that increase the maximum X-ray
intensity by as much as 1 dex (e.g.\ ROSAT observations of the Vela SNR
by \cite{boc94}) will not help our model fit the observations.  This is
because the brightest parts of the X-ray sky only cover a small
fraction of the sky, while in our model the brighter regions are large
scale features.

\placefigure{fig6}

\placefigure{fig7}

Another commonly computed parameter is the hardness ratio, which is
defined (for the two X-ray bands computed in this paper) as (I$_{\rm
0.70~keV}$ - I$_{\rm 0.22~keV}$)/(I$_{\rm 0.22~keV}$ + I$_{\rm
0.70~keV}$).  The hardness ratio is shown at each position with respect
to latitude in Figures 6a -- e and with respect to the column density
of \ion{H}{1} in Figures 7a -- e, and the hardness ratio of the strips
of the Wisconsin data that we have used previously are shown in Figures
6f and 7f.  In the plots from our model, we have added a small value to
the denominator of the ordinate to allow for cases where there is no
X-ray intensity in either band (for cases with no XRB).

The X-ray intensities appear to be a good discriminator when comparing
models to data, but the hardness ratios appear to be an even more
sensitive indicator of the success or failure of a calculation to
reproduce the data.  As we show, the emission from the models appears
to be significantly softer than those of the data, with the exception
of Position 5.  Of the bubbles, Bubble 3 has the largest fraction of
lines of sight with a distinctly positive hardness ratio. The overall
hardness of the X-ray spectrum at Position 5 is caused by a large
column of cold, dense gas within 60 -- 70 pc of the position that
prevents softer X-rays from reaching the center of the cold gas.  The
low temperature of Bubble 1 (compared with the average temperature in
Bubbles 2 and 3) creates the softness of the local gas in Bubble 1 (the
dashed line in Figure 6a), while only isolated pockets of the hottest
gas create maxima in the hardness plots.  For example, from the
position in Bubble 2 the hot gas in Bubble 3 subtends the latitude
range {\it b} = -60\arcdeg R -- -15\arcdeg R, and only the hottest gas
in the rightmost lobe of Bubble 3 (at {\it b} = -35\arcdeg R) has a
hardness ratio $>$ -0.50 (see Figure 6b).  As another example, the
hardest X-ray emission seen from Bubble 2 comes at an angle near {\it
b} = 0\arcdeg L, where the 0.70 keV emission is from Bubble 2 itself.
The dip near the midplane of the 0.70 keV band is also apparent in the
hardness ratio in Bubbles 1, 2, and 3 (only at {\it b} = 0\arcdeg L in
Fig.\ 6b).   As displayed in Figure 5, including the XRB (which has an
unattenuated hardness ratio = -0.50) hardens the resulting X-ray
spectrum at high latitudes from Bubble 1, and Positions 4 and 5 (see
Figs.\ 6a, 6d, and 6e).

Despite the small angular scale dip in the 0.70 keV intensity near the
midplane,  the emission usually hardens near the midplane (Figure 6),
which agrees with a well-know observational result, i.e.\ that all-sky
X-ray maps become less pole-dominated as the photon energy is increased
(see McCammon et al.\ 1983).  Since latitudes near the midplane are
associated with larger \hc, we have searched for a correlation between
hardness ratio of the X-ray intensities (without the XRB) and
\hc\ (Figure 7), and this is only strongly evident from Positions 4 and
5 (Figures 7d and 7e).  Adding the XRB does not change this result
much; specifically, the hardness ratio at low \hc\ sightlines is raised
from near -1.0 to the XRB value of -0.5.

\subsection{Anticorrelation Between H I and X-Ray Emission:  Models
vs.\ Observation}

The model data displayed as a function of latitude shows that there is
a complex mix of correlation and anticorrelation between the \hc\ and
I$_{\rm x}$ (after the extragalactic background has been added).   In
addition, the plots in Figures 4a and 4b (for Bubble 1) of X-ray
intensity in either band demonstrate that the intensity can appear {\it
correlated} with \hc\ rather than anticorrelated, at least on scales of
tens of degrees.  Specifically, the peak in \hc\ at {\it b} = 0\arcdeg
L for Bubble 1 coincides with a large scale rise in the X-ray intensity
in both bands.  However on a smaller angular scale, anticorrelation
does exist, e.g.\ the \hc\ maximum is coincident with the local minimum
in intensity (in both X-ray bands) at {\it b} = 0\arcdeg L in Figs.\ 4a
-- c.  In this subsection, we shall use Kendall's $\tau$ to quantify
the correlation between the \hc\ and I$_{\rm x}$ in each X-ray band.

A combination of two competing effects creates the combination of
correlation and anticorrelation between X-ray intensity and
\hc\ mentioned above.  First, dense regions of cold gas prevent hot gas
from expanding as quickly as would otherwise be the case, allowing the
hot gas to maintain its temperature; this generates a correlation
between X-ray intensity and \hc.  Another source of correlation is that
a shock moving into denser gas generates a higher emission measure
plasma than a shock moving into a lower density region.  Alternatively,
regions of dense gas also absorb non-local contributions to the X-ray
intensity, which improves the anticorrelation.  From the strip survey
data of model \hc, I$_{\rm 0.22~keV}$, and I$_{\rm 0.70~keV}$ in Figure
4, it appears that the former effect is somewhat more dominant for the
positions within bubbles, but both are apparent.

\placefigure{fig8}

For illustrative purposes, we have plotted the X-ray intensity
(including the larger XRB value in the 0.70 keV band) against the
\ion{H}{1} column density (see Figure 8) for Positions 3 and 4 in both
X-ray bands, as well as the observational data from the DL and
Wisconsin surveys.  Since some of the 0.22 keV model data are
coincident with the C band data in the plot, we have put the
observational data in separate panels from the model data to assist in
the comparison.  Comparing the model data from each position shows that
typical X-ray intensity in Figure 8 as observed from within Bubble 3 is
much larger than from Position 4.  The X-ray intensities are so weak
from Position 4 that some of the simulated data are not plotted in the
figure.  The observational data has a much narrower distribution (in
both coordinates, but particularly in the X-ray distribution) than any
of the model data.  However, there is a subsample of the 0.22 keV model
data that seems to match well with the entire C band strip, even to the
constant X-ray intensity feature that extends from log \hc\ $\sim$ 20.5
-- 22.0.  Even though this is coincidental, it may suggest that the
observations could be fit by a version of this model without too many
modifications.  No such coincidence is found between the M$_1$ band
data and the 0.70 keV band data; the model data, even for Bubble 3 ---
the most like the observational data (see Figure 8c), are roughly 1 dex
too faint.

The frequency of correlations or anticorrelations between the column
density of \ion{H}{1} and the X-ray intensity can be quantitatively
discussed by computing the non-parametric correlation Kendall's $\tau$
(e.g., \cite{pre89}) between \hc\ and I$_{\rm x}$.  This statistic
yields a value of -1 if both datasets are completely anticorrelated, 1
if they are completely correlated, and 0 if there is no correlation.

\placetable{tbl6}

From Kendall's $\tau$, one can compute the probability of obtaining a specific $\tau$ from uncorrelated
datasets (the null hypothesis).  In Table 6, we list both Kendall's $\tau$ and this probability for the complete \hc, I$_{\rm 0.22~keV}$, and
I$_{\rm 0.70~keV}$ model datasets for each of the five positions.
Since N$_{\rm cool}(b)$ is not a well-observed quantity, we have used
our computed values for \hc\ in this comparison.  
In addition, we have computed $\tau$ for subsets of the data
that are separated by the same angular scale, for as many subsets 
as the angular scale would allow (2 subsets of 360 data points separated
by 1\fdg0 degree, 3 subsets of 240 points separated by 1\fdg5 degrees,
etc.).  We did this for each subset with angular separations 
between 0\fdg5 and 10\fdg0 
in half-degree increments, and find that all of the $\tau$'s remain roughly 
constant as angular separation is varied, although the range in $\tau$ increased
as the separation increased (and the number of data points in the subsets
is decreased).  After deleting the data from the DL
\hc\ strip at the same latitudes where data had been removed from the
Wisconsin survey, we have computed the Kendall's $\tau$ of the strips
from the observational dataset.  These values are placed in columns 2
and 6 in the row labeled ``W" in Table 6.

The Kendall's $\tau$ reveals a distinct difference in the correlation
of the X-ray intensity and \hc\ for each of the two X-ray bands without
an XRB (column 2 for the 0.22 keV band and column 6 for the 0.70 keV
band in Table 6).  This difference is explained both by the different
mean free paths of 0.22 keV and 0.70 keV photons, and by the spatial
distribution of the hottest gas.  The three positions within bubbles of
hot gas have X-ray intensities in the 0.22 keV band that are either
positively correlated or uncorrelated (i.e., the probability that 
the specific $\tau$ could be from completely uncorrelated data is 
greater than 30\%) with \hc, while Position 4 (in warm gas) and Position 
5 (in cold gas) have distributions that are somewhat anticorrelated.  This situation is
reversed when the 0.70 keV intensity data are compared to the simulated
\hc\ data; the positions within bubbles are more anticorrelated than
for Positions 4 and 5, which show a significant correlation.  This
occurs because the 0.70 keV intensity is dominated by the hottest gas,
which rarely exists at high latitudes for positions outside of a
bubble, since the hottest gas is contained close to the midplane.  The
largest anticorrelation of \hc\ and 0.70 keV intensity is in Bubble 3;
one contribution to this anticorrelation is the 0.70 keV dip near {\it
b} = -90\arcdeg\ that coincides with a dense finger of cold gas that
separates two lobes in Bubble 3 (Figs.\ 1 and 4i).

The anticorrelation from the observational data, as described by the
Kendall's $\tau$ in the last row of Table 6, is stronger than that from
any of the positions we have chosen from the simulation, and is much
stronger than any in the 0.22 keV band.  The large difference between
the observed and model anticorrelations in the C band is most likely
caused by the shape and orientation of the bubbles that we have
analyzed.  Since none of Bubbles 1, 2, or 3 is larger in vertical
extent than in the midplane, as the Local Bubble is thought to be
(from the pole-dominated emission in Be, B, and C bands), the
anticorrelation of I$_{\rm 0.22~keV}$ with \hc\ in our model is
understandably smaller than the observational values.  None of the
model data has a similar decrease in the anticorrelation when the strip
from the Wisconsin C band is replaced by the strip from the M$_1$ band
in the comparison with the DL \hc\ strip, but this is mainly caused by
the lack of a good fit for the models in the softer band.  However, the
anticorrelation between the observational \hc\ and M$_1$ data is most
nearly matched by Bubble 3.

In all situations, additional X-ray emission from an XRB causes the
distributions to be more anticorrelated, with a result that all of the
$\tau$'s are negative for the largest values of E$_0$ used in each band
(columns 4 and 10 in Table 6).  This occurs despite the significantly
positive values of $\tau$ from positions 4 and 5 at 0.70 keV without
the XRB (column 6).  The most anticorrelated distribution of the model
data is between I$_{\rm 0.22~keV}$ + E$_0$ and \hc\ (column 4) at
Position 5, caused by the very large \hc\ at all {\it b} (minimum log
\hc\ $\sim$ 20.5) and the large effective cross-section at low X-ray
energies.

To summarize this subsection, our model generates a combination of
correlation and anticorrelations between \hc\ and the intensity in both
of the soft X-ray bands that we have computed.  However, for observers
within bubbles of hot gas there is a statistical tendency for
anticorrelations to be favored, and this trend is strengthened by the
addition of an XRB.

\subsection{X-Ray Model Spectra}

\placefigure{fig9}

In this section, we will discuss simulated spectra computed at a
latitude of +90\arcdeg\ from Bubble 2 and Position 4.  These lines of
sight will show the differences between a line of sight that has many
zones of hot gas with a large range of temperatures, including some
very hot gas in Bubble 2, and a line of sight that passes through a
superbubble in which the hot gas has cooled to a relatively low
temperature.  The model spectra (see Figure 9) were computed with a
high spectral resolution, with E/$\Delta$E $\sim$ 500 for each band
(435 for the mean energy in the 0.22 keV band, and 530 in the 0.70 keV
band).

\placetable{tbl7}

The computed spectra are displayed in Figure 9, as well as the spectra
convolved with energy resolutions, E/$\Delta$E (where $\Delta$E is the
FWHM of a Gaussian), of 30 and 100.  For the line of sight from Bubble
2, which contains some very hot gas, we find evidence for highly
ionized species (see Figs.\ 9c , 9i, and Table 7).  Such highly ionized
gas is not present for the line of sight through the cooled superbubble
(Figs.\ 9f, 9l, and Table 7), which passes through the relatively cool
gas (at T $\sim$ 5 $\times$ 10$^5$ K) of the superbubble at $x$ = 175
(see Figure 1a).

A single-temperature spectrum usually fits each of the model spectra.  
We have compared the highest resolution spectra shown (Figs.\ 9c, 9f, 9i,
and 9m) with single temperature spectra (assuming no intervening
absorption) described in \S4.3 by both matching continua by eye and
minimizing the usual Kolmogorov-Smirnov {\it D}-statistic. From this
analysis, we find that a Raymond-Smith plasma of log T = 6.30 is
similar to both spectra from Bubble 2 (Figs.\ 9c and 9i), while log T =
5.70 matches the 0.22 keV band spectrum from Position 4 (Figure 9f) and
log T = 5.95 does so for the 0.70 keV band spectrum (Figure 9l).  In
each of the simulated spectra plotted in Figure 9, we have added an XRB
spectrum with two characteristics: 1) an E$^{-1}$ slope (with no
emission lines), and 2) that the sum of intensities in each band is
equal to the XRB intensity that we estimated in \S 4.3 (the larger of
the two values in the 0.70 keV band).  We have convolved each of the
single temperature spectra with an energy resolution of E/$\Delta$E =
100, and plotted the result (the dotted line) in Figs.\ 9b, 9e, 9h, and
9k.  From these single temperature plots, one sees that the 0.70 keV
band spectra from Bubble 2 is not well fit by a single temperature
(Fig.\ 9h).  This results agrees with one from the K-S comparison, that
the {\it D}-statistic was minimized at two temperatures (log T = 6.25
{\it and} 6.45).  Despite the indication that absorption in the softer
band might have caused the multi-temperature ``fit" in the 0.70 keV
band (while this is not so in the 0.22 keV band), this is not true
because there is no gas at the higher temperature beyond Bubble 2
itself (at {\it b} = +90\arcdeg).

The spectra at different energy resolutions demonstrate the need for
high resolution X-ray spectra.  From the spectra in Figure 9, we show
that line identification can be accomplished only with an energy
resolution greater than between 30 and 100, while the detection of the
jagged continuum (caused by atomic line edges, e.g.\ 0.666 keV and
0.740 keV in Figure 9h) requires an E/$\Delta$E of at least 300.  Some
of the current or planned X-ray missions include the capability to
conduct high resolution spectroscopy (e.g., DXS, LEXSA, and XMM), with
E/$\Delta$E of about 100.

\section{Concluding Remarks and Directions for Future Work}

The calculations of a variety of model \ion{H}{1} and X-ray properties
has provided several insights into the nature of our interstellar
medium.  For the \ion{H}{1}, the observed scale height is reproduced in
the models, as is the general shape of \ion{H}{1} with latitude (for
Bubble 3), but the model \hc\ is too low at several high latitude
locations.  From within hot bubbles, the median \hc\ is similar to that
observed from the Sun, even though both values are below the mean
\hc\ through the disk (as an initial condition, the mean \hc\ in the
simulations of RB was taken to be equal to the value at the Solar
circle).  This difference, which in the model has the median value of
\hc\ a factor of 2.5 -- 6 lower than the mean, arises because
\ion{H}{1} has a low filling factor and there are relatively few
discrete \ion{H}{1} structures along a line of sight out of the disk.

Our calculations of model X-ray intensities show that the emission in
the two soft X-ray bands is dominated by hot gas that is close to the
observing position at 0.22 keV ($\lesssim$ 0.5 kpc).  At 0.70 keV, the
contribution from beyond the local bubble can dominate the emission,
although the non-local contribution varies widely with latitude; nearly
all the disk emission originated from $\lesssim$ 2 kpc.  Whereas
substantial variation occurs in the X-ray intensity (roughly a factor
of five or six in the Wisconsin C band and M$_1$ band data), the
variations in the calculated intensities are even greater, which points
out a shortcoming in the model.

An examination of the relationship between \hc\ and X-ray intensity
shows a complex combination of correlations and anticorrelations.
Aside from individual examples of each, we examined whether the two
observables are generally correlated or anticorrelated in a statistical
sense and sought to quantify the issue by calculating the nonparametric
correlation Kendall's $\tau$.  In the 0.22 keV band the \hc\ and the
X-ray intensity (without the XRB) are either uncorrelated or slightly
correlated from positions in bubbles of hot gas, while they are
anticorrelated from the positions of warm or cold gas, and this trend
is reversed in the 0.70 keV band (again without the XRB).  The
explanation for these trends is a combination of a short mean free path
of 0.22 keV photons and the proximity to the midplane of the hottest
gas that dominates the 0.70 keV emission.  However, as an XRB is added
to the X-ray intensity, the two simulated data sets become increasingly
anticorrelated.   The Kendall's $\tau$ for the strip of Wisconsin C
band data when compared with the DL \hc\ strip scan showed a very large
anticorrelation ($\tau$ = -0.6), that we feel is caused more by the
larger vertical extent of the Local Bubble in the Galaxy than any flaw
within our model.  This anticorrelation is reduced when the M$_1$ band
is compared with the \hc\ observational data ($\tau$ = -0.15), which is
similar to the anticorrelation between the model \hc\ and I$_{\rm
0.70~keV}$ data for Bubble 3.

We computed some spectra based upon the simulation, and found that
despite various contributions along the lines of sight a one
temperature Raymond-Smith spectrum often provided a satisfactory fit.
Multiple temperature fits were required along certain lines of sight,
although the ability to recognize a multi-temperature medium requires
good signal-to-noise spectra with an energy resolution of E/$\Delta$E
$\gtrsim$ 100.  This analysis argues for high energy resolution X-ray
spectrometers, since an energy resolution of E/$\Delta$E $\gtrsim$ 30
-- 100 to resolve many of the spectral lines.

The failure to reproduce certain important \ion{H}{1} and X-ray
observations has pointed out a shortcoming of the models.  For the
\ion{H}{1}, we find that the minimum \hc, which occurs at high
latitude, is too low as seen from all five locations.  In most, but not
all locations, the calculated variation of brightness temperature is
too great (e.g., the beam-to-beam fluctuations of the observations are
less than of the model).  In the X-ray band, the most pronounced
shortcomings of the models are that the X-ray surface brightness is too
low, and the range of X-ray variation is too great along the strip scans as
is the range of X-ray hardness ratios.  We consider whether these
shortcomings would be alleviated by varying some of the model
parameters (e.g., the supernova heating rate) or by adding physics
neglected in the existing model (e.g., magnetic fields).

A hotter, more pervasive model of the ISM would solve many of the X-ray
deficiencies of this model:  it would narrow the range of the X-ray
intensity strip scans, by increasing the intensity of the lowest values
in the 0.22 keV band, and it would raise all of the intensities in the
0.70 keV band (that would in turn increase hardness ratios).  An
increase in the volume filling factor of hot gas would occur if the
true supernova heating rate was a bit larger than we had assumed (this
number is probably unknown by a factor of three).  However, increasing
the hot gas filling factor would probably reduce the volume filling
factor of the neutral gas (see model F in RB), thereby reducing the
high latitude \hc, which is already too low.

The addition of magnetic fields to the simulations holds the
possibility of solving the greatest shortcomings of the model.
Magnetic fields can help confine hot gas in bubbles, making it more
difficult for such bubbles to break out of the disk.  This additional
confinement should lead to higher temperatures, densities, and
pressures in the bubbles and larger local emission measures, which
would raise the X-ray intensities of the models.  If breakout is less
common, then the overlying \ion{H}{1} layer will have fewer ``holes",
so there would be a reduction in the number low values of \hc\ at high
latitudes.  We plan to investigate whether these expectations are borne
out by quantitative calculations.  The significance of magnetic fields
on our models will be investigated in a future set of calculations,
which will incorporate the powerful diagnostics provided by the X-ray
and \ion{H}{1} observations.

\acknowledgments{The authors would like to thank John Raymond at CfA
for the use of the X-ray emission code, Mike Norman of the Laboratory
of Computational Astrophysics for providing the hydrocode ZEUS, Jay
Lockman for providing the DL \hc\ FITS file, Wilt Sanders for providing
FITS files of the Wisconsin survey C band and M$_1$ band data, and Dave
Davis for calculating the ROSAT conversions in Table 3.  Additionally,
we gratefully acknowledge David Burrows (the referee) for many helpful
suggestions from which the final version has benefited greatly.}

\clearpage

\begin{deluxetable}{ccccccc}
\tablewidth{0pt}  
\tablecaption{Local Characteristics of Positions \label{tbl1}}
\tablehead{
\colhead{Position \#} & \colhead{(x, z)} &  \colhead{height} & \colhead{area}
&\colhead{$\langle$n$\rangle$} & \colhead{$\langle$T$\rangle$} & \colhead{P$_{\rm med}$}\nl
&& (pc) & (kpc$^2$) & (cm$^{-3}$) & (K) &  (cm$^{-3}$ K)
}
\startdata
1    &  \phn55, 200 & \phn10 & 0.0188 & 6.435E-3          & 4.864E+5 & 3210 \nl
2    &  100, 224    & 230    & 0.0705 & 3.617E-3          & 1.464E+6 & 3060 \nl
3    &  135, 195    & \phn60 & 0.0382 & 6.788E-3          & 1.525E+6 & 7540 \nl
4    &  175, 200    & \phn10 & 0.0707 & 0.03552\phn       & 4.733E+4 & \phn630 \nl
5    &  \phn88, 200 & \phn10 & 0.0057 & 5.365\phn\phn\phn & 802.3\phn\phn\phn & 1420\nl
\enddata
\end{deluxetable}

\begin{deluxetable}{cccccccc} 
\tablewidth{0pt}
\tablecaption{Variations of N$_{\rm H~I}$\tablenotemark{a} \label{tbl2}}
\tablehead{
&&\colhead{low} && \colhead{high}\nl
\colhead {Pos.}  & \colhead{  min}   & \colhead{quartile}   & \colhead{median} 
& \colhead{quartile} & \colhead{max} & $\langle N_{\rm H~I} \rangle$ & \colhead{$\delta_N$} 
}
\startdata
1 &   0.0225 & 0.313 & \phn1.046 & \phn3.358 & 248.2 & \phn6.678 & 3.059\nl
2 &   0.0122 & 0.409 & \phn1.345 & \phn6.849 & \phn81.2 & \phn6.605 & 1.838\nl
3 &   0.0395 & 1.212 & \phn3.566 & \phn8.187 & 140.6 & \phn9.372 & 1.892\nl
4 &   0.0\phn\phn\phn & 0.103 & \phn0.801 & \phn4.111 & 231.5 & \phn6.961 & 2.832\nl
5 &   2.8431 & 7.160 & 13.502 & 20.415 & 235.3 & 16.663 & 1.142\nl
\tablevspace{8pt}
DL &  0.79\phn\phn  & 1.970 & 3.58\phn & \phn6.14\phn & 117.7 & 10.29\phn & 1.947 \nl
\enddata
\tablenotetext{a}{In units of 10$^{20}$ cm$^{-2}$}
\end{deluxetable}

\begin{deluxetable}{cccccc} 
\tablewidth{0pt}
\tablecaption{Conversion of Intensity Units to Counts per Second \label{tbl3}}
\tablehead{
\colhead{Energy (keV)} & \colhead{Wisconsin\tablenotemark{a}} & \colhead{ROSAT band} & \colhead{log T (keV)}& \colhead{log N$_{\rm H I}$ (cm$^{-2}$)} & \colhead{ROSAT\tablenotemark{b}}
} 
\startdata
0.22  & 2.76E5 & R2 & 6.0 & $\ldots$ &  1.8\phn\nl 
&&&    6.0 &  20.0 &  1.2\phn \nl
0.70  & 2.01E4 & R4 & 6.0 & $\ldots$ &  0.30\nl  
&&& 6.0 & 20.0 & 0.28 \nl
\enddata
\tablenotetext{a}{To units of cts s$^{-1}$}
\tablenotetext{b}{To units of cts arcmin$^{-2}$ s$^{-1}$}
\end{deluxetable}

\begin{deluxetable}{cccccccc} 
\tablewidth{0pt}
\tablecaption{Variations of I$_{0.22~keV}$\tablenotemark{a}~~with E$_0$ = 4.8E-5 \label{tbl4}}
\tablehead{
&&\colhead{low} && \colhead{high}\nl
\colhead {Pos.}  & \colhead{  min}   & \colhead{quartile}   & \colhead{median} 
& \colhead{quartile} & \colhead{max} & $\langle I \rangle$ &\colhead{$\delta_I$} }\startdata
1  &  2.85E-5\phn & 5.43E-5 & 9.46E-5 & 1.86E-4  & 1.44E-3 & 1.29E-4 & 0.880 \nl
2  &  1.73E-5\phn & 5.93E-5 & 9.18E-5 & 1.89E-4  & 1.10E-3 & 1.46E-4 & 1.024 \nl
3  &  1.80E-5\phn & 7.61E-5 & 1.37E-4 & 3.57E-4  & 9.75E-4 & 2.40E-4 & 0.905 \nl
4  &  9.38E-12 & 1.46E-5 & 3.46E-5 & 4.99E-5  & 3.45E-4 & 4.27E-5 & 1.006 \nl
5  &  1.05E-14 & 1.70E-8 & 1.50E-7 & 9.83E-7  & 4.22E-6 & 6.12E-7 & 1.447 \nl
\tablevspace{8pt}
C &   1.91E-4\phn & 3.68E-4\phn & 4.97E-4 & 7.34E-4 & 9.85E-4 & 5.30E-4 & 0.402\nl 
\enddata
\tablenotetext{a} {In units of 2.078 $\times$ 10$^{-12}$ ergs cm$^{-2}$ s$^{-1}$ arcmin$^{-2}$, see Table 3 for conversion to cts s$^{-1}$ arcmin$^{-2}$.}
\end{deluxetable}

\begin{deluxetable}{cccccccc} 
\tablewidth{0pt}
\tablecaption{Variations of I$_{0.70~keV}$\tablenotemark{a}~~with E$_0$ = 1.8E-5 \label{tbl5}}
\tablehead{
&&\colhead{low} && \colhead{high}\nl
\colhead {Pos.}  & \colhead{  min}   & \colhead{quartile}   & \colhead{median} 
& \colhead{quartile} & \colhead{max} & $\langle I \rangle$ & \colhead{$\delta_I$} }
\startdata
1\phn\phn\phn    &  8.80E-7 & 1.68E-5 & 2.01E-5 & 4.98E-5 & 4.48E-4 & 4.41E-5 & 1.219 \nl
2\phn\phn\phn    &  1.23E-5 & 3.02E-5 & 5.49E-5 & 8.58E-5 & 5.08E-4 & 7.62E-5 & 0.881 \nl
3\phn\phn\phn    &  2.33E-5 & 4.37E-5 & 6.33E-5 & 1.18E-4 & 4.00E-4 & 9.92E-5 & 0.803 \nl
4\phn\phn\phn    &  4.26E-7 & 1.71E-5 & 1.78E-5 & 2.68E-5 & 3.88E-4 & 4.14E-5 & 1.502 \nl
5\phn\phn\phn    &  5.61E-7 & 9.61E-6 & 1.39E-5 & 2.43E-5 & 1.69E-4 & 2.21E-5 & 1.036 \nl
\tablevspace{8pt}
M$_1$ & 3.22E-4\phn & 8.69E-4 & 1.03E-3 & 1.21E-3 & 1.714E-3 & 1.04E-3 & 0.240\nl
\enddata
\tablenotetext{a} {In units of 2.078 $\times$ 10$^{-12}$ ergs cm$^{-2}$ s$^{-1}$ arcmin$^{-2}$,
see Table 3 for conversion to cts s$^{-1}$ arcmin$^{-2}$.}
\end{deluxetable}

\begin{deluxetable}{ccccccccccc} 
\tablewidth{0pt}
\tablecaption{Correlation of N$_{\rm H~I}$ and I --- Kendall's $\tau$   \label{tbl6}}
\tablehead{
& \multispan2{\hfil 0.22 keV\hfil}& \multispan2{\hfil 0.22 keV + E$_0$\hfil}& \multispan2{\hfil 0.70 keV\hfil }& \multispan2{\hfil 0.70 keV + E$_{0,1.2}$\hfil }& \multispan2{\hfil 0.70 keV + E$_{0,1.8}$\hfil }\nl 
& \multispan2{\hrulefill}& \multispan2{\hrulefill}& \multispan2{\hrulefill }& \multispan2{\hrulefill }& \multispan2{\hrulefill}\nl
Pos.  &   $\tau$   &   Prob. &   
$\tau$   &   Prob. &   
$\tau$   &   Prob. &   
$\tau$   &   Prob. &   
$\tau$   &   Prob. \nl
(1) & (2) & (3) & (4) & (5) & (6) & (7) & (8) & (9) & (10) & (11) }
\startdata
1    & ~0.084  & 7.31E-4         & -0.071  &  4.22E-3       & ~0.072  &  4.07E-3         & -0.091 & 2.49E-4\phn         & -0.136 & 5.14E-8\phn \nl
2    & ~0.021  & 0.407\phn\phn   & -0.099  &  6.71E-5       & -0.078  &  1.75E-3         & -0.128 & 2.61E-7\phn         & -0.148 & 2.66E-9\phn \nl
3    & ~0.009  & 0.705\phn\phn   & -0.041  &  0.097\phn\phn & -0.128  &  2.89E-7         & -0.202 & 4.99E-16            & -0.231 & 1.49E-20 \nl
4    & -0.111  & 8.96E-6         & -0.438  &  0.000\phn\phn & ~0.286  &  0.000\phn\phn   & ~0.056 & 0.024\phn\phn\phn   & -0.010 & 0.674\phn\phn\phn \nl
5    & -0.047  & 0.059\phn\phn   & -0.577  &  0.000\phn\phn & ~0.370  &  0.000\phn\phn   & -0.078 & 1.73E-3\phn         & -0.145 & 6.31E-9\phn \nl
\tablevspace{8pt}
W    & -0.588  & 0.000\phn\phn      &         &          & -0.161  &  7.37E-6 \nl      
\enddata
\end{deluxetable}

\begin{deluxetable}{cccc} 
\tablewidth{0pt}
\tablecaption{Spectral Lines Seen in Figure 9 \label{tbl7}}
\tablehead{
\colhead{Position} & \colhead{X-ray Band (keV)}& \colhead{Energy (keV)} & \colhead{Contributing Ion(s)}
}
\startdata
2 & 0.22 & 0.161 & \ion{Fe}{14}  \nl 
        && 0.178 & \ion{Si}{8}, \ion{Fe}{15} \nl
        && 0.209 & \ion{Fe}{13}, \ion{Fe}{14}, \ion{Fe}{15} \nl
        && 0.251 & \ion{Si}{10}, \ion{Si}{11}, \ion{S}{9}, \ion{Ar}{9} \nl
2 & 0.70 & 0.56\phn & \ion{O}{7} \nl 
        && 0.57\phn & \ion{O}{7} \nl 
        && 0.65\phn & \ion{O}{8} \nl
        && 0.73\phn & \ion{Fe}{12} \nl 
        && 0.83\phn & \ion{Fe}{12} \nl
4 & 0.22 & 0.156 & \ion{Si}{7}, \ion{Fe}{12}, \ion{Mg}{8} \nl
        && 0.164 & \ion{Ne}{7}, \ion{Mg}{8} \nl
        && 0.170 & \ion{Si}{7}, \ion{Si}{8} \nl
        && 0.202 -- 0.204 & \ion{S}{7}, \ion{Si}{7}, \ion{Si}{8}, \ion{Si}{9}, \ion{Mg}{7}, \ion{Ne}{8}  \enddata
\end{deluxetable}

\eject

\figcaption{Grey scale images of gas temperature from a numerical
hydrodynamical simulation (see RB).  
a) This image of gas temperature shows the locations of the five viewing positions, and displays some sample lines of sight at different latitudes, {\it b}.  This figure also shows along the bottom and left axes the physical dimension of the grid, with sample lines of sight (5 kpc long) 
demonstrating straight paths in this physical space. 
b) This image shows all the gas above 300,000 K, which we use 
as the lower temperature limit of X-ray emitting gas.  None of the
sightlines from any of the positions will cross the hot gas 
at the extreme top and bottom of the figure, but this gas will not contribute 
significantly to the X-ray intensity, because the 
extremely low density leads to a low emission measure from these zones.   
c) This image displays the gas temperature of ``local" regions 
of a similar temperature to each of the viewing positions. 
\label{fig1}} 

\figcaption{Comparison of a slice of \hc\ observational data
and strip scans generated by the model.  The dotted line is
the data from the model and the solid line is a strip extracted from 
observational data (DL) that goes through both Galactic poles and 
longitudes of 90\arcdeg\ and 270\arcdeg.
The DL strip is shown starting with longitude 270\arcdeg\ 
at 0\arcdeg R.
\label{fig2}}

\figcaption{Histograms of the distribution of log \hc.  In panels a) -- e), we
show each set of 720 log \hc\ from the five positions that 
we analyzed; the panels are in the same order as Figure 2.   
In panel f), we show the distribution of the strip
extracted from the DL data.  Each of the histograms shown has
a bin size of 0.1 dex.  The lowest 1.5\% of the lines of 
sight (i.e., 11) from Position 4 have \hc\ = 0.0, and are not displayed 
in the histogram in panel d). 
\label{fig3}} 

\figcaption{The column density of both the cool gas and the cold gas (H I), 
and X-ray intensity as a function of latitude.  Panel a) -- c)
show strip scans for Bubble 1.  In the top panel, only model
data are shown, and the solid line represents log \hc\ while the dashed 
line is the column of all gas with T $<$ 300,000 K (N$_{\rm cool}$) 
within a distance of 5 kpc.  In the middle panel, we show X-ray data from the 0.22 keV band, and in the bottom panel, we show X-ray data from the 0.70 keV
band.  In the panels that show X-ray intensity, the solid line is the model intensity from within 5 kpc while the dashed line indicates the X-ray intensity including an extragalactic component, using one value for E$_0$ in the I$_{\rm
0.22~keV}$ plots and two values for E$_0$ in the I$_{\rm 0.70~keV}$
plots (see \S 4.3).  The dotted line shows the contribution from the local
bubbles (as defined in Figure 1c).   The strip scan extracted from 
the Wisconsin C and M$_1$ data are plotted as a dot-dash line (including
some anomalously low values where high brightness data had been omitted) 
in each of the  appropriate panels, with the data at longitude 270\arcdeg\ starting at 0\arcdeg R.  
Panels d) -- f) show strip scans for Bubble 2, with the same plots
as panels a) -- c).  
Panels g) -- i) show strip scans for Bubble 3, with the same plots
as panels a) -- c).  
Panels j) -- l) show strip scans for Position 4, with the same
plots as a) -- c) without dotted lines since there is no local 
bubble at this position.  
Panels m) -- o) show strip scans for Position 5, with the same
plots as a) -- c), again without dotted lines since there is no local 
bubble at this position.
\label{fig4}}

\figcaption{The cumulative X-ray intensity component for the 
incoming direction is
shown as a function of position along two lines of sight in the 0.22 keV
(solid line) and 0.70 keV (dashed line) bands.  The line of sight in panel
a) displays model data from Bubble 2, at {\it b} = -33\arcdeg R out of the midplane (in the direction of Bubble 3).  The model data in panel b) is 
from within the cool bubble at Position 4, along the midplane (at {\it b} = 1\arcdeg L).  By proceeding from the initial position, 5 kpc from the
observer (the right side of the diagram) leftward toward the observer
(situated at {\it d} = 0), we see how the intensity increases through emission
(regions of negative slope) and decreases by absorption of cool gas
(lines of positive slope).  The flat regions, where the intensity changes
little over some distance along the line of sight, occurs where the gas is neither hot enough to contribute to the intensity nor dense enough to absorb it
significantly.
\label{fig5}}

\figcaption{The X-ray hardness ratio [here $\equiv$ (I$_{\rm 0.70~keV}$
- I$_{\rm 0.22~keV}$)/(I$_{\rm 0.22~keV}$ + I$_{\rm 0.70~keV}$)]
vs.\ latitude.  Panels a) -- e) display the
model data from Bubbles 1, 2, and 3, and Position 4 and 5, respectively.
The dot-dashed line in panel f) shows the hardness ratio of the Wisconsin C and
M$_1$ bands (without the omitted high-brightness data), where each band 
has been scaled to our intensity units.  For the 
panels showing model data, the solid line displays the
intensities in each band out to 5 kpc without an XRB, the dotted line
(only in panels a -- c) shows the ratio from the gas in the local bubble,
and the dashed line shows the ratio for the intensities after
adding an XRB, where we used the brighter XRB added to the
I$_{\rm 0.70~keV}$ data.  The solid line at a hardness of zero shows where
the dominance of the softer band gives way to that of the harder band.
\label{fig6}}

\figcaption{The X-ray hardness ratio vs.\ \hc.  Panels a) -- e) display the
model data from Bubbles 1, 2, and 3, and Positions 4 and 5, respectively.
Panel f) shows the hardness ratio of the Wisconsin C and
M$_1$ bands (without the high-brightness data). Note that only the 
two positions near cool gas (Positions 4 and 5) have a strong 
dependence of X-ray hardness on \hc.  The solid line at zero has the same significance as in Figure 6.
\label{fig7}}

\figcaption{The X-ray intensities as a function of log \hc\ for all lines of sight for Bubble 3, Position 4 and the observational data.  Model
data are plotted by open pentagons and observational data are filled
squares.  The panels a) and b) show log I$_{\rm 0.22~keV}$ in Bubble 3
and Position 4, respectively, as panels c) and d) do for the log I$_{\rm 0.70~keV}$ model data.  The observational data are placed in the panels (not labeled) in the lower left and upper right to assist the reader in 
comparing the distributions along both axes.
\label{fig8}}

\figcaption{Simulated X-ray spectra with an XRB at different resolutions for 2 lines of sight.
The three resolutions correspond to E/$\Delta$E = 30, 100, and $\sim$ 500 
(the unconvolved data) for $b$ = +90\arcdeg\ from Bubble 2 at 0.22 keV in
panels a) -- c), at 0.70 keV in panels g) -- i), for $b$ = +90\arcdeg\ from
Position 4 at 0.22 keV in panels d) -- f), and at 0.70 keV in panels j) -- l). 
The XRB has a slope of E$^{-1}$ and no emission lines.  The dotted 
line in the E/$\Delta$E = 100 panels is a single
temperature spectrum with the XRB added convolved to have 
this energy resolution; this line does not appear in panel k)
because it is so close to the model data already plotted.
\label{fig9}}

\end{document}